\documentclass[conference]{IEEEtran}
\usepackage{blindtext, graphicx}
\usepackage[cmex10]{amsmath}
\usepackage{url}
\usepackage{csquotes}
\usepackage[english]{babel}
\usepackage[backend=bibtex,style=ieee,sorting=ynt]{biblatex}
\begin{document}
\title{A Modest Proposal for Open Market Risk Assessment to Solve the Cyber-Security Problem}

\author{\IEEEauthorblockN{Timothy J. O’Shea}
\IEEEauthorblockA{Bradley Department of Electrical\\and Computer Engineering\\
Virginia Tech, Arlington, VA\\
oshea@vt.edu}
\and
\IEEEauthorblockN{Adam Mondl}
\IEEEauthorblockA{Security Researcher\\
Austin, TX\\ 
launchpad.am@gmail.com}
\and
\IEEEauthorblockN{ T. Charles. Clancy}
\IEEEauthorblockA{Bradley Department of Electrical\\and Computer Engineering\\
Virginia Tech, Arlington, VA\\
tcc@vt.edu}
}

\maketitle

\begin{abstract}
We introduce a model for a market based economic system of cyber-risk valuation to correct fundamental problems of incentives within the information technology and information processing industries.  We assess the makeup of the current day marketplace, identify incentives, identify economic reasons for current failings, and explain how a market based risk valuation system could improve these incentives to form a secure and robust information marketplace for all consumers by providing visibility into open, consensus based risk pricing and allowing all parties to make well informed decisions.
\end{abstract}


\IEEEpeerreviewmaketitle

\section{Introduction}
The current state of cyber-security in the world is abysmal.  Most security experts have little to no confidence in the security of even their own computing systems or those employed by their organizations.  The largest nations in the world have difficulty protecting their most sensitive personnel files and secrets, the largest film studios can not protect their content or intellectual property, financial and private web services can not protect their sensitive customer data and there is no reason to think any of this will improve under the current incentive system.   

In this document we outline a brief summary of how a market based risk pricing system could correct the incentives involved in the cyber-security ecosystem and propose several potential institutions and market functions which could help to correct and align interests, risks and equities to help resolve the shortcomings and market failures of the cyber-economy as it exists today.  This market based valuation of risk solves the severe difficulties identified in recent work where a single insurer entity goes to great lengths to conduct risk assessment, and remains vastly under-informed. \cite{biener2015insurability} \cite{mukhopadhyay2013cyber}

\section{Motivation: Incentives in the Current System}

\begin{figure*}[ht!]
  \centering
      \includegraphics[width=1.0\textwidth]{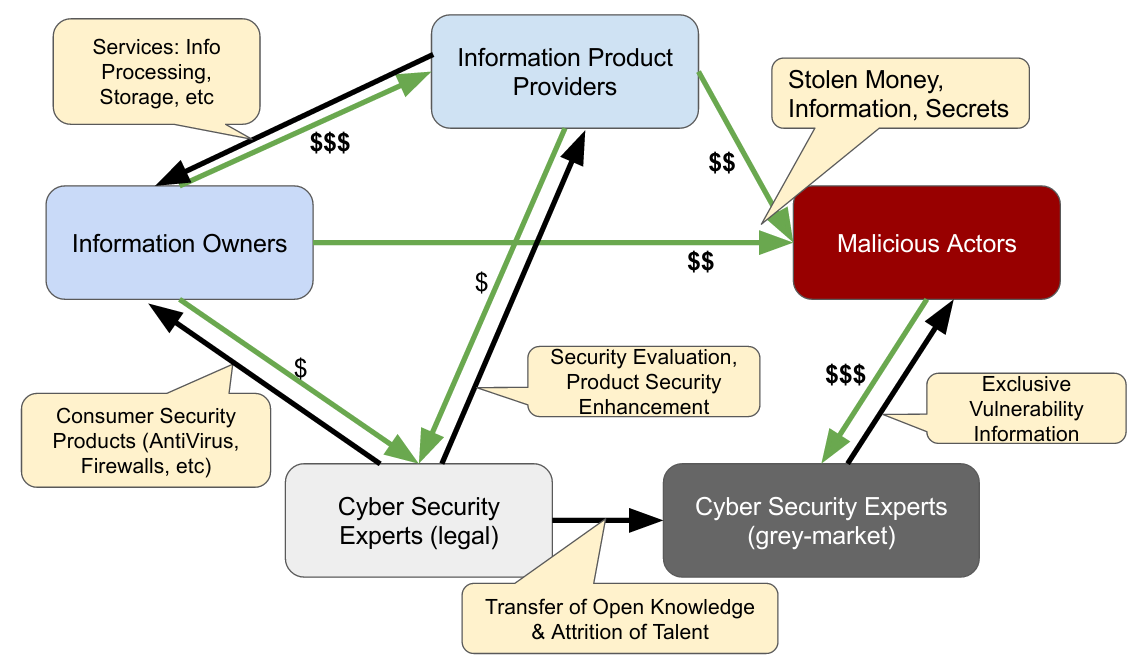}
        \caption{A simplified diagram of the economic cycles present in the current day cyber-security ecosystem demonstrating the misalignment of incentives }
    \label{fig:basis}
\end{figure*}
 
To understand why the system operates the way it does currently, we must explain the incentive system as it exists today.  To do this we attempt to reduce a complex marketplace for cyber security into several distinct classes of motivations, capabilities, and equities.  These are outlined below.

\textbf{Information Product Providers}: produce software apps, hardware platforms and web-based applications that we use everyday to communicate, store information, process information, and generally fill our information access and manipulation needs every day.   In many cases these are low margin industries where price competition for large consumer markets is key.  The market right now incentivizes these information product providers to provide the best functionality, at the lowest cost, and fastest delivery to potential buyers who may be interested.   These information product providers genuinely would love to provide security for their customers, but in general they can not afford to dedicate sufficient time, money, and effort into developing their products around security as a central function without sufficient long term customer demand to drive this cost benefit equation.   If a product succeeds in the market place, they can afford to attempt to bolt-on security later, but unfortunately that is generally not how such systems work, and a ground up approach to security is necessary.

In summary information product providers:

\begin{itemize}
  \item Seek to fill the customer's need for functionality, low cost, fastest delivery
  \item Seek to provide security for customers but often don't know how or can not invest sufficiently upfront into this focus.
\end{itemize}

\textbf{Information Owners, Product Consumers}: This constitutes all of us.  Each of us participates in email, text messaging, calling, storing documents, filing taxes, online banking, online investment, and any number of information operations which allow us to function efficiently every day through electronic means.   To accomplish this we must choose hardware to operate on, be they computers, smartphones, or tablets, we must choose an operating system, a browser, a series of applications, websites and online services with whom we entrust our information and privacy to accomplish these tasks.  Often the decisions on which platforms, software, or services to use are centered around cost decisions, familiarity, ease-of-use, performance criteria, product features, visual user interface features, or other such factors.  Some people are heavily security conscious when making such choices, but the vast majority of users either rank security and privacy lower in their priority list, or more likely are not informed enough to be able to properly evaluate security features and risks in terms of their decisions.
In summary information owners

\begin{itemize}
 \item Seek to achieve some information task either storage or processing
 \item Seek to minimize their cost and time spent accomplishing these tasks
 \item Seek ease of use, familiarity and cost of time spent learning new tasks
 \item Seek security and privacy but often don’t know how to properly evaluate or value it
\end{itemize}

\textbf{Cyber Security Experts (within the legal system)}: 
are generally seeking to fill the need of either information owners or information product providers for some for-profit motivation.   Since there is a large demand from some information owners or product providers who are security conscious, such as banks and tax businesses that have a large interest in this, there is potentially a sizable incentive here, but often buyers and sellers are uninformed and unable to properly value risks, solutions, or providers, and often unable to pay premiums for high quality solutions when they are identifiable.   In general cyber security experts seeking to operate clearly within the legal system choose to operate in this role.
In summary a legal-market cyber security expert:

\begin{itemize}
\item Seek to identify security products to maximize profit from customers (information owner and product providers)
\item In general seek to identify risks and provide suggestions for product development practices and security features
\item Are widely varied in technical and risk assessment abilities and skill levels, but are generally not easily valued properly by the market
\end{itemize}

\textbf{Cyber Security Experts ( Grey-Market )}: are security experts possessing various valuable skills who seek to maximize the potential return on their work.  Due to the often vast discrepancy between selling security products, bug information or receiving bug-bounties from information product owners and that of selling undisclosed vulnerability information and proof-of-concepts on anonymous markets or to brokers of vulnerabilities who deal with all variety of vulnerability consumers, the latter often becomes an attractive option, offering significantly higher potential for returns than pursuing so-called responsible disclosure or public disclosure.   
In summary grey-market cyber security experts:
\begin{itemize}
\item Have a unique ability to discover and leverage unknown vulnerabilities
\item Seek to maximize their return from these vulnerabilities
\item Are not concerned with operating within legal and moral boundaries
\item Are widely varied in technical and risk assessment abilities and skill levels, but can be clearly valued based on the potential impact of the vulnerability for sale.
\end{itemize}

\textbf{Malicious Actors (illegal activity)} are generally those who seek to leverage vulnerability information or social engineering for their own monetary, political, informational, criminal or national gains.  In general these are formed by groups with a strong motivation to participate in one of these acts to increase power, money, or control in some community.   These groups are often well funded and can be involved in highly profitable enterprises where a specific high-impact enabling vulnerability provides an extremely high value proposition to them.  For instance a single vulnerability for the proper system or software may provide a group access to millions of sensitive records, information, or otherwise which may enable significant amounts of follow-on profitable illegal activity.   If used properly, a vulnerability may be used numerous times without disclosure or diminishing value over time.  Upon disclosure, value may remain high if there is little incentive, motivation or knowledge to patch effected systems quickly.
In summary malicious:

\begin{itemize}
\item Seek to steal, modify, disrupt, remove, or otherwise tamper with private information
\item Place a large incentive on cyber security experts to provide undisclosed vulnerabilities
\item Pose risks to both information owners’ private information, and information product providers’ product integrity and security
\end{itemize}

 \section{The Solution: Correcting Incentives}

 \begin{figure*}[ht!]
  \centering
      \includegraphics[width=1.0\textwidth]{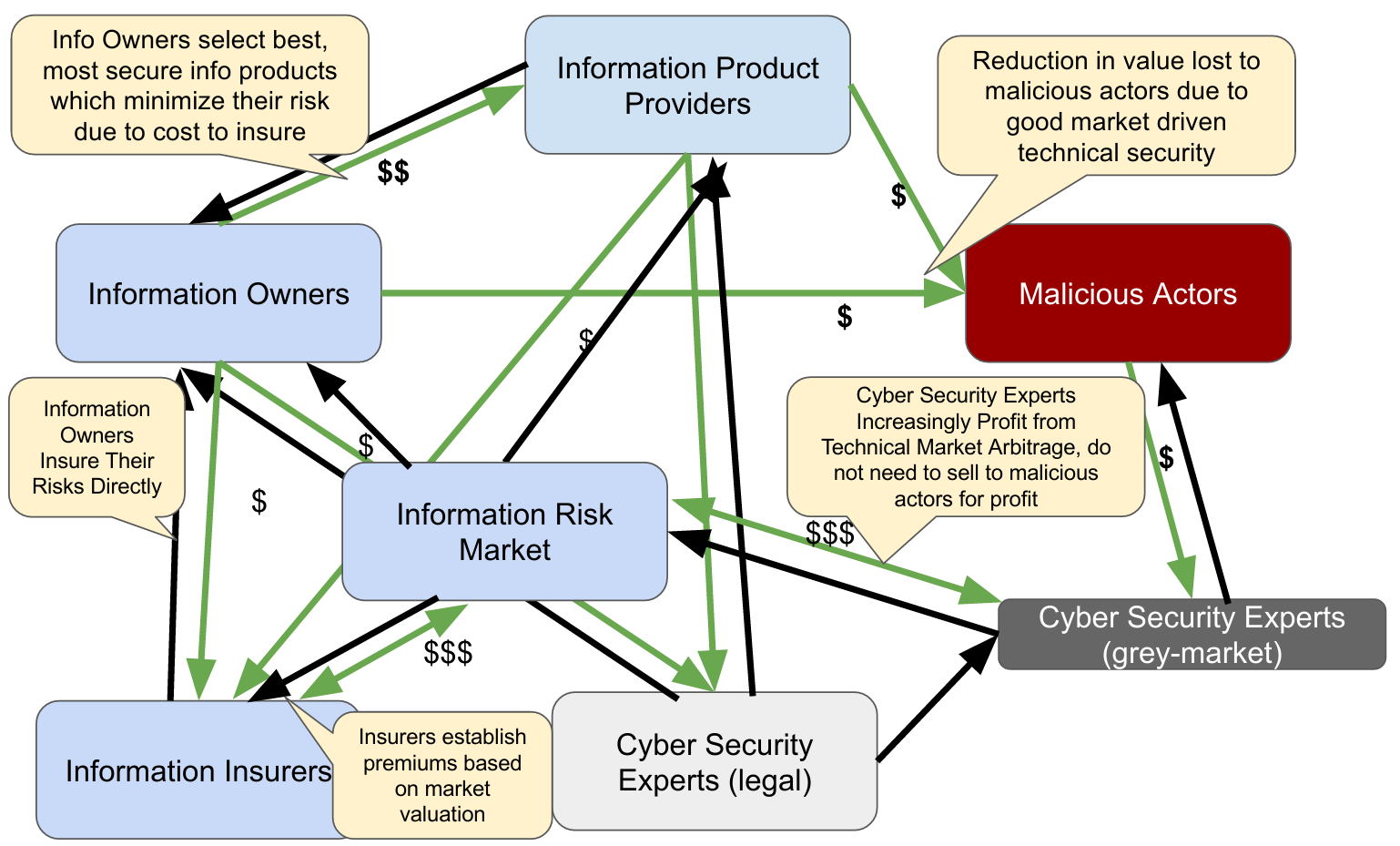}
        \caption{Our updated version of our cyber-security ecosystem cycle including the proposed institutions of market based risk valuations in information products as well as information security insurance providers operating with these collaborative valuations}
    \label{fig:system2}
\end{figure*}

 Research into solving problems into the cyber security dilemma has been principally focused on technical problems and technical countermeasures.  This has resulted in a number of promising techniques which do improve security and have been valuable.   Stack guards, memory protections, role based access control, containers, ISA-enhancements, formal-methods, and other technical solutions have done much to advance the field, but adoption and deployment of best practice methods is abysmally slow such that much of this research from a practical user’s perspective is irrelevant or extremely slow to reach market.   This class of technical methods unfortunately tends to live well below the visibility of most information product providers when making design and software decisions, let alone the information owner's’ level.  It is unfortunately the norm that decision makers for such information products are either unqualified, apathetic, or unwilling to undertake the extra effort or expense of making the most prudent technologic choices to promote security.  It's hard to fault them to some extent since there is little customer pressure to do so, users and information owners generally don’t know any better and have little if any visibility into such choices made by product providers.
 
To fundamentally address these issues and enhance security for information owners and information product providers in this system we need to better inform consumers and product providers by providing crystal clear consensus valuations of both risks and costs associated with information systems and technical measures to secure them.

This is no trivial task, as real knowledge in such risk valuation is extremely rare and contentious among numerous parties.   Unfortunately we have very little accountability or incentive at the moment to ensure that “experts” in the field are held accountable to their recommendations, endorsements, and so called wisdom in the field.  Previous work has suggested cyber-risk insurance \cite{gordon2003framework} \cite{mukhopadhyay2005insurance}, and this is available today, but due to the poor ability of a single entity to accurately evaluate risk without a market consensus it must be priced too high for general use to make sense. \cite{briys1991reliability}

Many of these same qualities could be said for insight into valuation of equities, risk assets, and other financial products.   Generally the solution to pricing such assets has fallen to free market systems as our best current approach.   Allowing those with confidence in future valuations of assets to stake positions which benefit them proportionally to their confidence and investment in such positions is one relatively well understood way to allow all parties to contribute and benefit from, given good information, a robust free market system.

By establishing better quantified risks for information products, we allow insurers to begin to offer informed risk products to consumers, allowing for risk policies which pay out some premium upon the loss of privacy or compromise of specific information.   Given that this is a more predictable event, consumers will be influenced by policy premium and product providers will be influenced to best engineer their systems to minimize premiums and incentivize customers to their product offering.   By allowing a 3rd party insurance outlet, we avoid the issue of information product risk compensation negligence, in which information product companies simply can not tolerate a default on information risk of their product, and disappear and cease to provide service upon an insecure product being once compromised.
We believe that one system that might significantly improve the functioning of the current cyber security ecosystem is that of introducing a market system for establishing confidence in certain information products.   By doing so we allow a number of critical effects which drastically change the ecosystem, outlined below.

\textbf{Information Owners}: Selection of products which the market generally endorses as secure products, without need for understanding deep technical details.  Selecting products which deliver needed capabilities with minimum risk policy premiums and select premium sizes in accordance with their personal risk tolerance.

\textbf{Information Product Developers}: Gain an incentive to seek real technical means and organizational practices to significantly bolster their market-evaluated risk and security measures. Wish to provide technical means which minimize risk, and minimize risk policy pricing for risk instruments of their customers for using their products.

\textbf{Security Experts}: Perhaps the most significant effect, security experts with actual impactful security enhancing techniques, or security experts having discovered significantly impactful vulnerabilities in narrow or widely used systems can benefit from them by trading positions which will reap major financial rewards from a public release of such a vulnerability, and then releasing them publicly and openly. This is an option far superior to selling them privately and without disclosure to malicious actors, and has the potential to dwarf the incentive for doing so through the collective risk valuation of a product.

\textbf{Information Security Insurance Providers}: Lastly we introduce the concept of an information security insurance providers, whom the information owner relies on to advise (potentially through premium pricing) on which information product owners’ products to use, while providing some level of monetary risk compensation for the leak and release of the user’s sensitive information.  This is an entity which largely does not exist right now, but with better quantified market-agreement on risk levels of various providers and technical means could serve an important role in guiding consumers to safety. My personal belief is that such a system is yearning to exist in the current time, but without better market agreed upon risk and loss valuations, any company doing so would either be foolish or need to charge absurd premiums to serve such a role.

\section{Cyber-Risk Arbitrage}
The primary method by which we propose to correct the incentive system is that of introducing a market mechanism of cyber-risk speculation or arbitrage.   That is, introducing market-priced risk instruments which allow consumers to easily assess the level of risk associated with each product, allowing them or privacy insurance companies to take steps to price and insure privacy risk at the level appropriate for information and individuals, and by providing a quantitative financial benefit to information service providers for building to and investing in risk mitigation strategies through the returns of and their confidence in their risk backed instruments.

By allowing cyber security experts to directly transfer technical knowledge into market pricing knowledge through arbitrage and speculation, and possible public information release, we offer a greater outlet than ever before for pricing assets in line with the reality of their security.   We begin to remove grey-market and low-knowledge distortions which plague the whole ecosystem by hiding exploits from the general public, and by helping to inform general decision making quantitatively without requiring heavy technical expertise.

\section{Reducing Grey-Market Activity and Malicious Actors}

Economists have long discussed the pros and cons of prohibition, resulting in grey markets for illicit goods.  Typical current and past grey markets consist of alcohol, drugs, weapons, and prostitution.   The morality of these items aside, the economic evidence is often that prohibitions are not effective, do not work, and largely empower and make wealthy those willing to accept the risks of engaging in criminal enterprise, and those who they pay to make this possible.   

Grey markets are grey for two reasons:

\begin{itemize}
 \item The sale or activities in question are illegal and thus carry a large risk upon them.
 \item Consequently such markets generally operate with lower visibility, consumer knowledge of competitive pricing distributions, and with few informed product option to make safe choices on their own, making them grey.
\end{itemize}

While there has been discussion of creating a formal grey-market through the use of export regulations in so called “cyber-munitions” lists to formalize the first of these two reasons, it is largely not yet applicable and bears no domestic implications and faces a huge barrier of technical implementation feasibility.
The second however is exactly what we have before us.   A market in which very few actors are setting prices in closed transactions, the vast majority of people are uninformed as to the risks, technologies, capabilities, and ultimately the choices they must make to ensure their security.

Therefore, to reduce and improve the functionality of this market for security and privacy, we must do three things.

\begin{itemize}
\item Make security valuation an open process where multiple skilled valuation entities may clear to help price the market intelligently.
\item Allow consumers to make educated decisions when choosing security products based on market valuation of risk.
\item Provide incentive for technical experts to achieve maximum profit by assisting the legal market to best function for themselves, consumers and information product providers.
\end{itemize}

\section{Industry Path Forward}
The process for establishing such a market place is yet to be determined, this could take the place of a privately run market trading in potential disruption events, or a government facilitated market institution.   Legal or consume rating requirements could rapidly incentivize the adoption of such a system for consumers, and in general, market insurers and security experts should follow as they both begin to see and be able to measure and predict the value in various arbitrage scenarios.   Once this happens, this system should have a positive self-reinforcing effect, which tends to make the system stronger by ensuring all parties of aligned interests.

If this takes the form of a privately run market, we certainly see the incentive for being a market maker in an early established system.   For providing a market making service, there exists significant potential to profit off transactions in an early, and fundamentally-necessarily growing marketplace as it becomes larger and more liquid during the adoption curve.

\section{Conclusion}

The problem of cyber-security will largely always be an economic problem.  Until enough collective self-interest in secure products and services is aligned into a single functioning market entity, the incentive of security researchers to engage with malicious actors in the sale of cyber capabilities, whether through legal or illegal means, will continue to far outweigh the alternative.  

It is critical that for this market to function, risk-valuation both of information product providers and the software and configurations they use become a distributed and market-based system, allowing all experts to partake in pricing and imparting their knowledge through the practice of buying or selling risk instruments relating to products or companies of interest.  

Providing such an option allows the global cyber-security system to select computer security experts with the most successful risk valuation abilities and vulnerability knowledge to help contribute to security rather than to insecurity while potentially greatly increasing the incentive for providing such a security role.  It provides an important method for which both information owners and information product providers can look to market guidance for their choices in software, companies, and practices followed to minimize their own risk exposure and minimize insurance premiums.  

Lastly it serves to significantly increase the cost to malicious actors of funding and participating in cyber-crime by creating a functioning market system able to price in publicly disclosed information and create major price competition for the use of undisclosed security information.

\appendices

% \section*{Acknowledgment}
%The authors would like to thank Virginia Tech

\nocite{friedman1990free}
\nocite{ehrlich1972market}
\nocite{anderson2006economics}

\printbibliography

%\begin{IEEEbiography}[{\includegraphics[width=1in,height=1.25in,clip,keepaspe%ctratio]{picture}}]{John Doe}
%\blindtext
%\end{IEEEbiography}

\end{document}